\documentclass[aps,twocolumn,amsmath,amssymb,showpacs,amsfonts,prd,footinbib]{revtex4}
\usepackage{epsfig}
\usepackage{graphicx}
\usepackage{dcolumn}
\usepackage{bm}
\usepackage{amsmath, amsthm, amssymb}
\usepackage{color}

\newcommand{\bea}{\begin{eqnarray}}
\newcommand{\eea}{\end{eqnarray}}

\begin{document}
\title{On Scale Versus Conformal Symmetry in Turbulence}
\author{Yaron Oz}
\affiliation{Raymond and Beverly Sackler School of
Physics and Astronomy, Tel-Aviv University, Tel-Aviv 69978, Israel}
\date{\today}
\begin{abstract}
We consider  the statistical description of steady state fully developed incompressible fluid turbulence at the inertial range of scales in any
number of spatial dimensions. We show that turbulence statistics is scale but not conformally covariant, with the only possible 
exception being the direct enstrophy cascade in two space dimensions.
We argue that the same conclusions hold for compressible non-relativistic turbulence as well as for relativistic  turbulence. We discuss
the modification of our conclusions in the presence of vacuum expectation values of negative dimension operators.
We consider the issue of non-locality of the stress-energy tensor of inertial range turbulence field theory.

\end{abstract}


\maketitle


Fully developed incompressible fluid turbulence is largely considered as the most important unsolved problem of classical physics.
Most fluid motions
in nature at all scales are turbulent.
Aircraft motions, river flows, atmospheric phenomena, astrophysical flows and even blood flows
are some examples of set-ups where turbulent flows occur.
Despite centuries of research, we still lack
an analytical description and understanding of fluid flows in the non-linear regime.
Insights to turbulence hold a key to understanding the principles and
dynamics of  non-linear systems with a large number of strongly interacting degrees of freedom far from equilibrium.

One defines the inertial range to be the range of length scales $l \ll r \ll L$,  where the scales $l$ and $L$ are
determined by the viscosity and forcing, respectively.
Experimental and numerical data suggest that
turbulence at the inertial range of scales  reaches a steady state that exhibits statistical homogeneity and isotropy
and is characterized by universal scaling exponents that depend only on the number of space dimensions $d$ \cite{LL,Frisch} (for a proposal of
anomalous scaling setup at low Reynolds number  see \cite{V}).

The aim of this letter is to consider the question whether turbulence statistics realized by the universal scaling exponents can exhibit, as many critical systems do, not only scale symmetry
but rather conformal symmetry.  We will use exact scaling relations to show, in general,  that turbulence statistics is scale but not conformally covariant.
We will find one possible  exception, which is the direct enstrophy cascade in two space dimensions.
Furthermore, we will argue that the same conclusions hold for compressible non-relativistic turbulence as well as for
relativistic  turbulence. We will discuss how the presence of  vacuum expectation values of negative dimension operators (condensates)
can modify these conclusions.

Under a $d$-dimensional conformal transformation:
\begin{equation}
x^i \rightarrow  x'^i~~~ dx'^2 = \Omega^2 (x) dx^2 ,~~i=1,...,d \  ,
\label{conformal}
\end{equation}
and
\begin{equation}
\frac{\partial x'^{i}}{\partial x ^{j}} = \Omega(x) R^{i}_{j}(x),~~~R^{i}_{k}(x)R^{k}_{j}(x) 
=\delta^{i}_{j} \ , \label{R}
\end{equation}
thus $R^{i}_{j}(x) \in SO(d)$. Raising and lowering indices is done with Kronecker delta $\delta_{ij}, \delta^{ij}$.
Conformal transformations rescale lengths non-uniformly, while preserving the angles between vectors.
The conformal group includes the Euclidean group that consists of translations, rotations and dilatations as well
as special conformal transformations.
Special conformal transformations are composed of an inversion $x^i \rightarrow  \frac{x^i}{x^2}$ followed by a translation $x^i\rightarrow x^i+a^i$
and by a second inversion. They
take the form :
\begin{equation}
x^{i} \rightarrow  x'^{i} = \frac{x^{i} + a^i x^2}{1+ 2 a\cdot x + a^2 x^2}  \ .
\end{equation}

Consider first the case of inertial range incompressible fluid turbulence that describes fluid flows with a low Mach number. 
We are interested in the statistics of the turbulent velocity vector field $v^i$ that satisfies the incompressibility 
condition $\partial_iv^i=0$, and hence in the correlation functions :
\begin{equation}
\langle v^{i_1}(x_1)\cdot\cdot\cdot v^{i_n}(x_n) \rangle \ , \label{cor}
\end{equation}
where  the separation between points $x_{ij}$ in (\ref{cor}) is in the inertial range. 
We will work in the limit  $l \rightarrow 0$  and $x_{ij}$ fixed in which 
experimental and numerical data suggest that the correlation functions (\ref{cor}) are finite. The correlation functions have also a 
a finite limit when $x_{ij}\rightarrow 0$, after taking the limit $l \rightarrow 0$, which allows us to define composite
operators $v^{i_1}(x)v^{i_2}(x)\cdot\cdot\cdot v^{i_k}(x)$ \cite{OPE}.
The dimension of these operators is $k$-times the dimension of $v^i$.

Consider Kolmogorov's law \cite{Kolmogorov} :
\begin{equation}
\langle v^{ij}(x_1) v^k(x_2) \rangle = \epsilon (\delta^{ik}x_{12}^j +\delta^{jk}x_{12}^i - \frac{2}{d}\delta^{ij}x_{12}^k) \ ,
\label{exact}
\end{equation}
where  $\epsilon$ is the mean rate of energy dissipation due to viscosity, and  
$v^{ij}(x)$  is a traceless symmetric 2-tensor of $SO(d)$
\begin{equation}
v^{ij}(\vec{x}) =   v^i(x) v^j(x) - \frac{v^2(x)}{d} \delta^{ij}   \ . \label{v2irred}
\end{equation}
Kolmogorov's law (\ref{exact}) is an exact relation in statistical turbulence, and thus 
provides a consistency check on any proposal for such a description.
In the following  we will show that  (\ref{exact}) is not compatible with conformal symmetry. 

The mean rate of energy dissipation $\epsilon$ in
(\ref{exact}) is a dimensionful constant that depends on the forcing scale $L$. This is generally case with the coefficients 
of turbulence structure functions. Thus, the scale versus conformal symmetry that we consider corresponds to the universal 
scaling exponents of the structure functions.

The operators of conformal field theories are classified as primary or descendants.
A conformal primary operator of dimension $\Delta$  in an irreducible  representation $R^I_J$ of $SO(d)$ transforms under a conformal
transformation (\ref{conformal}) as :
\begin{equation}
O_I(x) \rightarrow  O'_I(x')  =  \Omega(x)^{-\Delta} R^{J}_I(x) O_J (x) \ . \label{primary}
\end{equation}
Descendants  are derivatives of the primary operators and their transformation can be deduced from (\ref{R}) and
(\ref{primary}).
Consider two primary operators $O_I$ and $O_J$ with scaling dimensions $\Delta_I$ and $\Delta_J$, respectively.
Scale (dilatation) symmetry $x^i \rightarrow \lambda x^i, \lambda = const$, restricts the form of their two-point function:
\begin{equation}
\langle O_I(x_1) O_J(x_2) \rangle = \frac{c_{IJ}(x_{12})}{x_{12}^{\Delta_I+\Delta_J}} \  ,
\label{2pt}
\end{equation}
where $c_{IJ}$ is dimensionless.
Special conformal  symmetry sets further restrictions.
Under a conformal transformation (\ref{conformal}) we have :
\begin{equation}
x_{12}^2 \rightarrow  {x'}_{12}^{2}  = \Omega(x_1)\Omega(x_2) x_{12}^2 \ .
\label{res1}
\end{equation}
Using (\ref{primary}), (\ref{2pt}) and (\ref{res1}) one gets that conformal covariance of two-point function (\ref{2pt}) requires 
$\Delta_I = \Delta_J$. This is clearly  not the case in (\ref{exact}), since the dimension of $v^i$ is not equal to the dimension
of $v^{ij}$.
 In fact the orthogonality theorem requires that the two operators transform in the same irreducible $SO(d)$ representation 
 \cite{Ferrara:1972xq} and
 \begin{equation}
c_{IJ}(x_{12}) \rightarrow  R^K_I(x_1)R^L_J(x_2) c_{KL}(x_{12}) \ . \label{res2}
\end{equation}
However, $v^i$ and  $v^{ij}$ transform in different irreducible representations  of $SO(d)$.

It is possible that the operators $v^i$ or $v^{ij}$ (or both) are not primary operators.
Indeed, consider the divergence free velocity vector $v^i$.
The two-point function of a conserved spin one operator $J^i, \partial_iJ^i=0$ of  dimension $\Delta$ reads :
\begin{equation}
\langle J^{i}(\vec{x}_1) J^{j} (\vec{x}_2)   \rangle  =   \frac{\delta^{ij} + a \frac{x^i x^j}{x^2} }{x_{12}^{2 \Delta}},~~~a = \frac{2 \Delta}{d -2 \Delta -1} \ . \label{2Vgen}
\end{equation}
The requirement to be a conformal primary (\ref{primary}) implies that  $a=-2$ in (\ref{2Vgen}), hence $\Delta= d-1$.
Thus, for  $v^i$  to be a conformal primary operator its dimension should have been $d-1$. However, its dimension is 
experimentally close to its Kolmogorov (K41) linear scaling dimension \cite{Kolmogorov}
$\Delta_{K41}[v^i] = - \frac{1}{3}$ (in inverse length units) \cite{Benzi}.

Since $v^i$ cannot be a primary operator let us  assume that it is a descendant of a primary operator. 
It is clearly not a descendant of $v^{ij}$,  $v^i \neq \partial_jv^{ij}$. Thus, the two-point function of $v^i$ and $v^{ij}$ can be nonzero only if they are 
both descendants of the same primary operator, and in such a case 
their dimensions should differ by an integer 
number. This is not  case:
Their K41 linear scaling dimensions differ by $-\frac{1}{3}$, while 
the actual experimental value may differ  slightly from this value  \cite{Benzi}.

As a second example consider the
exact scaling relation of incompressible fluid turbulence derived in \cite{Falkovich:2009mb}:
\begin{equation}
\langle v^i(x_1)p(x_1)v^2(x_2) \rangle =  C x_{12}^i  \ , \label{vp}
\end{equation}
where $C$ is a constant related to the mean rate of energy dissipation, and $p$ is the fluid pressure.
In this case the two operators are  $v^i(x) p(x)$ and $v^2(x)$ and their dimensions differ but not by  an integer.
Thus, the two-point function (\ref{vp}) is not conformally covariant.

Two-dimensional fluid turbulence is special since there are two cascades: the direct enstrophy (vorticity squared) cascade and the 
 inverse energy cascade.
 The analysis of the inverse energy cascade works as above and it is not conformally covariant (see, however, \cite{SLE}
for a numerical evidence for a confromal structure in isovorticity lines).
In the direct enstrophy cascade one can derive the exact scaling relation:
 \begin{equation}
\langle \omega(x_1)\omega(x_2)v^i(x_2) \rangle = C x_{12}^i \ , \label{direct}
\end{equation}
where $\omega = \epsilon^{ij}\partial_iv_j$ is the vorticity pseudoscalar 
and $C$ is a constant related to the mean rate of enstrophy dissipation.
In this case the two operators are  $\omega(x)$ and $\omega(x)v^i(x)$. Their dimensions may differ by one (the dimension of $v^i$ is -1 or close to it) and it is still possible
that they are both descendants of the same  primary operator.
This is the case considered by Polyakov  in \cite{Polyakov:1992er}.

Consider next compressible non-relativistic fluid flows, where the fluid density  $\rho(x)$  is not constant.
In this case, one can derive an exact scaling relation that takes the form \cite{Falkovich:2009mb}  :
\begin{equation}
\langle T^{0i}(x_1) T^{ij}(x_2) \rangle = \epsilon x_{12}^j \ , \label{T1}
\end{equation}
where we sum over the index $i$, $\epsilon$ is a constant related to the mean rate of energy dissipation and
\begin{equation}
T^{0i} = \rho v^i,~~~~T^{ij} = \rho v^iv^j + p \delta^{ij} \ , \label{r}
\end{equation}
satisfies the ideal  compressible fluid equation $\partial_{t}T^{0j} + \partial_{i}T^{ij} = 0$.
The two-point function (\ref{T1}) reduces to the Kolmogorov law (\ref{exact}) in the incompressible case $\rho=const$, when using
$\langle p(x_1)v^i(x_2) \rangle = 0$ which follows from incompressibility, and using isotropy to recast it in the manifestly isotropic form  (\ref{exact}). 
We have in  (\ref{T1})   two different operators that do not have dimensions 
that differ by an integer. Hence (\ref{T1})  cannot be a two-point function of a confomally invariant theory. 
 
 In two space dimensions there is an enstrophy cascade of compressible fluid flows similar
to the incompressible fluid one.
One can derive an exact scaling relation for such turbulent flows that takes the form \cite{Westernacher-Schneider:2015gfa}:
\begin{equation}
\langle \omega^j(x_1)\omega(x_2)\rangle = C x_{12}^j \ , \label{directcomp}
\end{equation}
where 
\begin{equation}
\omega^j = \epsilon^{ik}\partial_kT_i^j,~~~~\omega = \epsilon^{ik}\partial_kT_{0i} \ ,
\label{om}
\end{equation}
and  $C$ is a constant related to the mean rate of enstrophy dissipation.
The  scaling relation (\ref{directcomp}) has been checked numerically in \cite{Westernacher-Schneider:2015gfa,Westernacher-Schneider:2017snn}
and it reduces to (\ref{direct}) in the limit of non-relativistic fluid flows.
In this case the difference between the dimensions of the two operators which is the dimension of $v^i$
may be an integer  as in the incompressible limit, thus it is possible that it is conformally covariant.

Let us discuss now relativistic hydrodynamics defined by the conservation of a relativistic stress-energy
$T^{\mu\nu},\mu,\nu=0,...,d$, $\partial_{\mu}T^{\mu\nu} = 0$.
The stress-energy tensor  of an ideal  relativistic fluid reads:
\begin{equation}
T^{\mu\nu} = (\varepsilon + p)u^{\mu}u^{\nu} + p \eta^{\mu\nu} \ , \label{T}
\end{equation}
where $\varepsilon$ is the energy density, $p$ is the relativistic pressure, $u^{\mu} = (\gamma, \gamma\frac{v^i}{c})$, 
$\gamma = (1-\frac{v^2}{c^2})^{-\frac{1}{2}}$,
 is the relativistic
fluid velocity vector $u_{\mu}u^{\mu} = -1$ and $\eta^{\mu\nu} = diag[-,+,...,+]$ is the Minkowskian metric.
One can derive an exact scaling
\cite{Fouxon:2009rd} of the form (\ref{T1}) with (\ref{T}),
which reduces to the Kolmogorov law  (\ref{exact}) in the limit of non-relativistic fluid flow $v\ll c$.
The difference between the dimensions of the two operators in (\ref{T}) is the dimension of $v^i$, which is unlikley
to be an integer (it is not an integer in the non-relativistic limit). This suggests that also relativistic turbulence is not conformally
covariant. 

In two space dimensions there is also an enstrophy cascade of relativistic fluid flows similar
to the incompressible fluid one.
One can derive an exact scaling relation for such turbulent flows that takes the form (\ref{directcomp}), (\ref{om}) and (\ref{T}) \cite{Westernacher-Schneider:2015gfa},
which reduces to (\ref{direct}) in the limit of non-relativistic fluid flows.
In this case the difference between the dimensions of the two operators which is the dimension of $v^i$
may be an integer  as  in the non-relativistic limit, thus it is possible that it is conformally covariant.

Consider the K41 theory of turbulence \cite{Kolmogorov}. Since  K41 theory neglects intermittency, the dimension of a general composite
operator of the form $v^{i_1}(x)v^{i_2}(x)\cdot\cdot\cdot v^{i_k}(x)$  is $\frac{k}{3}$.
Similarly to the above analysis, we conclude that K41 theory is scale but not conformally covariant, by e.g. using the Kolmogorov law.
In \cite{Oz:2017ihc} we proposed a field theory of turbulence at the inertial range of scales
and derived the formula for anomalous scalings of the longitudinal structure functions  proposed in \cite{Eling:2015mxa}. The field theory is
based
on dressing the K41 mean field theory by  a conformal field theory of a gapless dilaton mode, and we related the intermittency to the boundary 
conformal anomaly coefficient. The discussion here suggests that we should include in the intermittency also the scale anomaly coefficients.

In our analysis we assumed that the vacuum expectation values of the turbulence field theory operators are zero. 
However, since we have operators $O_K$ with negative dimensions $\Delta_K < 0$ they may acquire expectation values \footnote{I thank A. Polyakov for pointing this out to me.} :
\begin{equation}
\langle O_K \rangle = c_K L^{-\Delta_K} \ ,
\label{cond}
\end{equation}
where $c_K$ are nonzero constants and $L$ is the infrared forcing scale. The expectation values (\ref{cond}) break spontaneously
conformal symmetry and modify the correlation functions of conformal field theories \cite{Zamolodchikov:1990bk}. For instance, the two-point correlation function (\ref{2pt}) will be modified due to such expectation values
by dominating terms of the form \cite{Polyakov:1992er} :
\begin{equation}
 \frac{c_{IJK}(x_{12})}{x_{12}^{\Delta_I+\Delta_J}} \left(\frac{L}{x_{12}}\right)^{|\Delta_K|} \ ,
\end{equation}
where $c_{IJK}$ is dimensionless. In such a case
one cannot use the orthogonality theorem \cite{Ferrara:1972xq}.
The issue whether such nonzero expectation values exist requires a study of the infrared boundary conditions and dynamics of the turbulent system.

Finally, let us make a comment about the stress-energy tensor of the field theory of inertial range incompressible fluid turbulence. A local stress-energy
tensor is a  dimension $d$ conserved 2-tensor of $SO(d)$. It  is  easy to see on dimensional grounds that we cannot make such an object 
from the velocity vector $v^i$ and the pressure $p$
in the energy cascades, hence the stress-energy tensor of a field theory description of turbulence will necessarily be non-local.
This is not in contradiction with field theory axioms that require the existence
of energy and momentum but do not require the existence of their densities.
In the enstrophy cascade in two space dimensions it is possible that such a local stress-energy tensor exists.

\section*{Acknowledgements}
I would like to thank G. Falkovich, V. Mukhanov, A. Polyakov and V. Yakhot for discussions.
This work  is supported in part by the I-CORE program of Planning and Budgeting Committee (grant number 1937/12), the US-Israel Binational Science Foundation, GIF and the ISF Center of Excellence.

\end{document}